\documentstyle[12pt,amsfonts]{article}
\newcommand{\bR}{{\mathbb R}}

\newcommand{\bC}{{\mathbb C}}

\newcommand{\kB}{{\cal B}}
\newcommand{\kH}{{\cal H}}
\newcommand{\kO}{{\cal O}}
\newcommand{\kQ}{{\cal Q}}

\newcommand{\imag}{{\Im{\mathrm m\,}}}
\newcommand{\dom}{{\mathrm{dom}}}
\newcommand{\ran}{{\mathrm{ran}}}
\newcommand{\sign}{{\mathrm{sgn}}}

\newtheorem{claim}{Claim}[section]
\newtheorem{theo}[claim]{Theorem}
\newtheorem{prop}[claim]{Proposition}
\newtheorem{lem}[claim]{Lemma}
\newtheorem{cor}[claim]{Corollary}

\newtheorem{rem}[claim]{Remark}

\begin{document}

\title{Potential approximations to $\delta'$: an inverse Klauder
phenomenon with norm-resolvent convergence}
\author{Pavel Exner,$^{a,b}$ Hagen Neidhardt,$^{c}$ and
Valentin A. Zagrebnov$^{d,e}$}
\date{}
\maketitle \maketitle

\begin{quote}
{\small \em a) Department of Theoretical Physics, NPI, Academy of
Sciences, CZ-25068 \v Re\v z \\
 b) Doppler Institute, Czech Technical University, B\v{r}ehov{\'a} 7,
CZ-11519 Prague, \\ \phantom{e)x}Czech Republic \\
 c) Weierstra{\ss}-Institut f\"ur Angewandte Analysis und Stochastik,
Mohrenstr. 39, \\ \phantom{e)x}D-10117 Berlin, Germany \\
 d) D\'{e}partment de Physique, Universit\'{e} de la
M\'{e}diterran\'{e}e (Aix-Marseille II)\\
 e) Centre de Physique Th\'{e}orique, CNRS, Luminy Case 907, F-13288
Marseille \\ \phantom{e)x}Cedex 9, France \\
 \rm \phantom{e)x}exner@ujf.cas.cz, neidhard@wias-berlin.de,
 zagrebnov@cpt.univ-mrs.fr}
\vspace{8mm}

\noindent {\small Abstract:} We show that there is a family
Schr\"odinger operators with scaled potentials which approximates
the $\delta'$-interaction Hamiltonian in the norm-resolvent sense.
This approximation, based on a formal scheme proposed by Cheon and
Shigehara, has nontrivial convergence properties which are in
several respects opposite to those of the Klauder phenomenon.
\end{quote}

\section{Introduction}
\setcounter{equation}{0}

Point interactions are often used for constructing solvable models
of quantum mechanical systems \cite{AGHH}. To judge a quality of
such models one has to be able, of course, to decide how well does
a point interaction approximate the ``actual'' interaction. In the
simplest case of a one-dimensional point interaction introduced
originally by Kronig and Penney \cite{KP} the answer is easy: the
appropriate Hamiltonian is a norm-resolvent limit of a family of
Schr\"odinger operators with squeezed potentials, which physically
means that a slow particle with a widely smeared wave packet
``sees'' only the mean value of a localized potential. The problem
is more complicated in dimension two and three where squeezed
potentials can also be used, however, with a renormalization such
that the limiting coupling is ``infinitely weak''. The idea
belongs to Friedman \cite{Fr}; a detailed discussion for a general
shape of the approximating potential can be found in \cite{AGHH}
together with description of other, nonlocal, approximations and
the corresponding bibliography. More generally, we have here an
important particular case of the question what is the ``right
Hamiltonian'' for a strongly singular perturbation of the Laplace
operator -- see \cite{NZ1, NZ2} and references therein.

A peculiarity of the one-dimensional situation is that not all
point interactions are of the $\delta$ type. This follows from the
standard construction of a point interaction \cite{BF, AGHH} which
relies on restriction of the free Hamiltonian to functions which
vanish in the vicinity of the interaction support, and a
consecutive construction of self-adjoint extensions of the
obtained symmetric operator. For a single center in dimension one
the latter has deficiency indices $(2,2)$ leading thus to a
four-parameter family of extensions. A subset of them usually
called $\delta'$ interactions was introduced in \cite{GHM}, the
whole family was later studied in \cite{GeK, Se1, GH} and
subsequent papers by other authors. In distinction to the usual
$\delta$ interactions the other extensions were constructed as
mathematical objects and the question about their physical meaning
arose naturally.

\v{S}eba \cite{Se2} was the first who addressed the question of
approximation of $\delta'$ Hamiltonians by those with ``regular''
interactions. He showed, in particular, that the name is
misleading because such Hamiltonians {\em cannot} be obtained
using families of scaled zero-mean potentials. At the same time he
demonstrated that the $\delta'$ interaction can be approximated in
a nonlocal way using a suitable family of rank-one operators with
a nontrivial coupling-constant renormalization. Later local
approximations were constructed \cite{Ca,CH} but they were not of
potential type since they involved first-derivative terms.

The question about the $\delta'$ interaction meaning became more
appealing when interesting physical properties of this coupling
were discovered. Specifically, it was shown that Wannier-Stark
systems with an array of $\delta'$ interactions have no absolutely
continuous spectrum \cite{AEL, Ex, MS} and even that the spectrum
is pure point for most values of the parameters \cite{ADE}. The
origin of this effect is the high-energy behaviour of the
$\delta'$ scattering, with the transmission amplitude vanishing as
$k\to\infty$. Such a behaviour can be approximated, up to a phase
factor, within a fixed finite interval of energies by small
complicated graph scatterers \cite{AEL}, and the qualitatively
same scattering picture, up to a series of resonances, was found
for a sphere with two halflines attached \cite{Ki}.

Until recently it was believed, however, that no potential-type
approximation to the $\delta'$ interaction existed. It came thus
as a surprise, when two years ago Cheon and Shigehara (CS)
constructed an approximation by means of a triple of $\delta$
interactions with the coupling constants scaled in a nonlinear way
as their distances tend to zero \cite{CS}. In distinction to the
situations mentioned above this renormalization leads to an
``infinitely strong'' coupling in the limit. The said authors
computed formally the limiting wave function and showed that it
obeyed the $\delta'$ boundary conditions \cite{AGHH}; they also
presented an alternative argument based on convergence of the
corresponding transfer matrices \cite{SMMC}.

It is natural to ask in which sense does the limit exist and
whether one can construct a similar approximation using regular
potentials. We shall answer the second question affirmatively and
show that the approximating families converge in a rather strong
topology, namely norm resolvent. A nontrivial character of the
approximation will be seen from the fact that we do not recover
the sought limit when the involved operators are expressed through
the respective quadratic forms, in particular, because the form
domain of the limiting operator is {\em larger} than those of the
approximating ones.

Such a disparity between the form domains reminds of the {\em
Klauder phenomenon} \cite{Kl, Si} where a singular perturbation is
switched off in the strong resolvent sense yielding an operator
different from the free one obtained as the formal limit by
putting the coupling constant equal to zero. Here the situation is
in several respects opposite. First of all, the coupling here is
not switched off as in \cite{Kl, Si} but rather becomes {\em
infinitely strong}, so it is not straightforward to identify the
formal limit. On the other hand, the larger form domain
corresponds to the {\em true} norm-resolvent limit. In addition,
the CS-approximation requires a subtle interplay of the coupling
constants. If we change this choice, we arrive at an operator the
form domain of which is {\em smaller} than those of the
approximants, namely to the Laplacian with Dirichlet decoupling at
the $\delta'$ interaction position.

Let us review briefly the contents of the paper. In the next
section we will formulate the approximation by triple $\delta$
interaction and examine it using the explicit form of the
operators involved. Then we combine this result with the known
squeezed-potential approximation of the $\delta$ interaction
\cite[Thm.~I.3.2.3]{AGHH} to show that a $\delta'$ can be
approximated by a family of potentials consisting of three
$a(\epsilon)$-spaced parts of a ``size'' $\epsilon$ which approach
each other as $\epsilon\to 0+$ and at the same time undergo a
CS-type scaling. Furthermore, we determine a squeezing rate which
yields a convergent approximation: it is sufficient that
$\epsilon\, a(\epsilon)^{-12}$ tends to zero. In Section~4 we
illustrate the mentioned nonstability of the approximation: if we
disbalance only slightly the $\epsilon$ dependence of the coupling
constants we get a family which converges in the norm-resolvent
sense to the Dirichlet decoupled Laplace operator on the line. To
keep things simple we do not strive for a maximum generality. We
restrict ourselves to the $\delta'$ case, because an extension to
the general four-parameter point interaction is easy to obtain by
adapting the scheme of \cite{SMMC}. We also do not ask about the
optimal rate between $\epsilon$ and $a(\epsilon)$ needed for the
convergence.


%
\section{Resolvent approach to the CS approximation}
\setcounter{equation}{0}

In the following we use the notations and definitions of
\cite{AGHH}. Let $H_0 = -\Delta$ be free one-dimensional
Schr\"odinger operator in the Hilbert space $ L^2(\bR)$. Its
resolvent is an integral operator with the kernel
\begin{equation} 
G_k(x\!-\!x') \equiv(-\Delta -k^2)^{-1}(x,x') :=
{\frac{i}{2k}}\, e^{ik|x-x'|}
\end{equation}
for any $\imag k> 0$ and $x,x'\in \bR$. The related function
\begin{equation} \label{signG}
\tilde G_k(x\!-\!x') := \sign(x\!-\!x')\, {\frac{i}{2k}}\,
e^{ik|x-x'|}
\end{equation}
allows us to express the resolvent for the
${\delta}'$-perturbation of $H_0$ centered at the point $y$ and
having the ``strength'' $\beta$, denoted by $\Xi_{\beta,y}$, in
the form  \cite[Sec.~I.4]{AGHH}:
\begin{equation} \label{Krein}
(\Xi_{\beta,y} - k^2)^{-1}(x,x') = G_k(x\!-\!x') -\,
\frac{2{\beta}k^2}{2-i{\beta}k}\: {\tilde G_k(x\!-\!y)} {\tilde
G_k(x'\!-\!y)} \,.
\end{equation}
Recall that $\Xi_{\beta,y}$ acts as $H_0$ away of $y$ and its
domain consists of those $f\in W^{2,2}\left( \bR \setminus \{ y\}
\right)$ which satisfy the boundary conditions
\begin{equation} \label{delta'bc}
\psi'(y+) =\psi'(y-) =:\psi'(y)\,, \quad \psi(y+)-\psi(y-)
=\beta\psi'(y)\,.
\end{equation}
Our first aim is to approximate the resolvent (\ref{Krein}) of
$\Xi_{\beta,y}$ by a family of those corresponding to the triple
$\delta$-perturbation of $H_0$ with the couplings ${\mathcal{A}}_a
= \{\alpha_j\}_{j=-1,0,1} = \{2\beta^{-1} \!-\! a^{-1}, \beta
a^{-2}, 2\beta^{-1} \!-\! a^{-1} \}$ localized at $Y_a =
\{y_j\}_{j=-1,0,1} = \{y-a,y,y+a\}$ for $a \geq 0$ letting $a
\rightarrow 0$. Denote this perturbed operator by $-
\Delta_{{\mathcal A}_a,Y_a}$ . Then by \cite[Sec.~II.2]{AGHH} the
corresponding resolvent has the kernel
\begin{equation} \label{3delta-res}
( -\Delta_{{\mathcal A}_a,Y_a}-k^2)^{-1}(x,x') = G_k(x\!-\!x') -\!
\sum_{j,j'= -1,0,1}[\Gamma_a(k)]_{jj'}^{-1}\, { G_k(x\!-\!y_j)\,
G_k(x'\!-\!y_{j'})}  \,,
\end{equation}
where $[\Gamma_a(k)]_{jj'} := \left\lbrack
{\alpha}_j^{-1}{\delta_{jj'}} + G_k(y_j\!-\!y_{j'})
\right\rbrack_{jj'}$ and $j,j' = -1,0,1$. In particular, for a
purely imaginary $k = i \kappa$, $\kappa > 0$, we get
\begin{equation} \label{Gamma}
\Gamma_a(i\kappa)= \frac{1}{2\kappa} \left(
\begin{array}{ccc}
 1 + u & w & w^2 \\
 w & 1 + v &  w \\
 w^2 &  w & 1 + u
\end{array} \right) \,.
\end{equation}
 where
\begin{equation} \label{uvw}
u := 2\beta\kappa a/(2a \!-\! \beta), \quad v:= 2\kappa a^2/\beta,
\quad w := e^{-\kappa a}.
\end{equation}
Let us look how the spectrum of the operators
$\{-\Delta_{{\mathcal A}_a,Y_a}\}_{a\geq0}$ behaves as  $a
\rightarrow 0$ for a fixed $\beta$. Since the perturbation in
(\ref{3delta-res}) is a rank three operator, $\sigma_{ess}(H_0) =
\sigma_{ac}(H_0) =[0, \infty)$ is not affected by the perturbation
and the point spectrum consists of at most three negative
eigenvalues, with the multiplicity taken into account
\cite[Sec.~8.3]{We}. Here we have:
\begin{prop} \label{spec approx}%
For small enough spacing $\,a\,$ the operator $-\Delta_{{\mathcal A}_a,Y_a}$
has at most one eigenvalue. This happens if and only if $\,\beta <
0$, and in that case
\begin{equation} 
\inf \sigma(-\Delta_{{\mathcal A}_a,Y_a}) = -\,{4\over {\beta}^2} + \kO(a)
\,.
\end{equation}
\end{prop}
{\em Proof:} Since the negative part of
$\sigma(-\Delta_{{\mathcal A}_a,Y_a})$ is the point spectrum determined by
zeros of $\det\Gamma_a(i\kappa)$ by \cite[Sec.~II.2]{AGHH} we
arrive at the equation
\begin{equation} 
(1\!+\!u\!-\!w^2) \left\lbrack (1\!+\!u)(1\!+\!v)-w^2(1\!-\!v)
\right\rbrack =0 \,,
\end{equation}
or
\begin{equation} \label{spec1}
 e^{-2\kappa a}=1+{2\beta\kappa a \over 2a \!-\! \beta}
\end{equation}
and
\begin{equation} \label{spec2}
 e^{-2\kappa a}=\left( 1+{2\beta\kappa a \over 2a \!-\!
 \beta}\right)\,
 {1+{2\kappa a^2} \beta^{-1} \over 1-{2\kappa a^2}\beta^{-1}} \,.
\end{equation}
Expanding the left- and right-hand sides of the last two equations
around $a=0$, one finds that only (\ref{spec1}) has a solution for
a sufficiently small $a > 0$ and that it equals
$$ \kappa(a) = -\frac{2}{\beta} + \kO(a).$$
Since $k = i\kappa$ corresponds to an isolated eigenvalue if and
only if $\imag k > 0$, the assertion follows readily. \hfill$\Box$
\vspace{2mm}

Proposition \ref{spec approx} also shows that if $\kappa >
-2/\beta$, $\beta \not= 0$, is fixed, then there is $a_0(\kappa) >
0$ such that $-\Delta_{{\mathcal A}_a,Y_a} + \kappa^2 > 0$  and
the resolvent $(-\Delta_{{\mathcal A}_a,Y_a} + \kappa^2)^{-1}$
exists for all $a \in (0,a_0(\kappa))$. We further note that the
operator $-\Delta_{{\mathcal A}_a,Y_a}$ admits a definition in the
sense of quadratic forms. Denoting by $\kQ_{{\mathcal
A}_a,Y_a}[\cdot,\cdot]$ this quadratic form one has
\begin{eqnarray}\label{quadratic}
\lefteqn{\kQ_{{\mathcal A}_a,Y_a}[u,v] = (u',v') +
\frac{\beta}{a^2}\, u(y)\overline{v(y)} }\\ & & \phantom{AAAAA} +
\left(\frac{2}{\beta} -
\frac{1}{a}\right)\left\{u(y+a)\overline{v(y+a)} +
u(y-a)\overline{v(y-a)}\right\}\nonumber
\end{eqnarray}
for $u,v \in \dom(\kQ_{{\mathcal A}_a,Y_a}) = W^{1,2}(\bR)$. When equipped
with the scalar product
\begin{equation} \label{quadratic1}
(u,v)_{\kQ_{{\mathcal A}_a,Y_a}} := \left(\sqrt{-\Delta_{{\mathcal A}_a,Y_a} +
\kappa^2}\;u,\sqrt{-\Delta_{{\mathcal A}_a,Y_a} + \kappa^2}\;v\right)\,,
\end{equation}
where $\kappa > -2/\beta$ and $a \in (0,a_0(\kappa))$, the domain
$\dom(\kQ_{{\mathcal A}_a,Y_a})$ becomes a Hilbert space. It is
important to note that the norm $\|\cdot\|_{\kQ_{{\mathcal
A}_a,Y_a}}$ arising from this scalar product is equivalent to the
norm of the Hilbert space $W^{1,2}(\bR)$.

Proposition \ref{spec approx} shows that up to an $\kO(a)$ error
the spectral properties of $-\Delta_{{\mathcal A}_a,Y_a}$ coincide
with those of $\Xi_{\beta,y}$. Next we compare the corresponding
resolvents.
\begin{theo} \label{delta approx}
Let $\kappa \not= -2/\beta$ and $\beta \not= 0$ be fixed. Then the
relation
\begin{equation} \label{resolv approx}
\lim_{a\rightarrow 0+}\, \left( -\Delta_{{\mathcal A}_a,Y_a} + \kappa^2
\right)^{-1}(x,x')= \left(\Xi_{\beta,y} + \kappa^2
\right)^{-1}(x,x')
\end{equation}
holds for any $x,x'\in \bR$. Consequently, $-\Delta_{{\mathcal A}_a,Y_a} \to
\Xi_{\beta,y}$ as $a\to 0+$ in the norm-resolvent sense.
\end{theo}
{\em Proof:} By virtue of (\ref{Krein}), to check (\ref{resolv
approx}) it is sufficient to compute the pointwise limit of the
second term at the right-hand side of (\ref{3delta-res}). Using
the notations introduced in the preceding proof, we obtain an
explicit expression for the inverse matrix in (\ref{3delta-res}):
\begin{eqnarray} \label{inverse}
 \lefteqn{
 [{\Gamma_a(i\kappa)}]^{-1} = \frac{2\kappa}
 {(w^2\!-\!1\!-\!u)[(1\!+\!u)(1\!+\!v)-w^2(1\!-\!v)]} } \\ && \nonumber \\
&&  \times\left(
\begin{array}{ccc}
 w^2-\!(1\!+\!u)(1\!+\!v) & -w(w^2\!-\!1\!-\!u) & w^2v \\
 -w(w^2\!-\!1\!-\!u) & (w^2\!+\!1\!+\!u)(w^2\!-\!1\!-\!u) &
 -w(w^2\!-\!1\!-\!u) \\
 w^2v &  -w(w^2\!-\!1\!-\!u) & w^2-\!(1\!+\!u)(1\!+\!v)
\end{array} \right) \,. \nonumber
\end{eqnarray}
Without loss of generality we may assume $y=0$. Suppose, for
instance, that $x,x'>a$, then the resolvent difference kernel is
obtained by sandwiching the above matrix between the vectors
$G(x),\, G(x')$, where
\begin{equation} \label{sandwich}
G(x):= \left( \begin{array}{c} G_{i\kappa}(x+a) \\ G_{i\kappa}(x)
\\ G_{i\kappa}(x-a) \end{array} \right) = {1\over 2\kappa}\, e^{-\kappa x}
\left( \begin{array}{c} w \\ 1 \\ w^{-1} \end{array} \right)\,,
\end{equation}
which yields the expression
\begin{equation} \label{dif}
\sum_{j,j' = -1,0,+1}[\Gamma_a(i\kappa)]^{-1}_{jj'}G(x - y_j)G(x -
y_{j'}) =  {1\over 4\kappa^2}\, e^{-\kappa x} \, e^{-\kappa x'} \,
{N\over D}
\end{equation}
with
\begin{equation}\label{D}
D = \frac{(w^2\!-\!1\!-\!u)[(1\!+\!u)(1\!+\!v)-w^2(1\!-\!v)]}
{2\kappa}
\end{equation}
and
\begin{equation}\label{N}
N = (w^2+w^{-2}) [w^2- (1\!+\!u)(1\!+\!v)] + 2w^2v +
(w^2\!-\!1\!-\!u) (u\!-\!1\!-\!w^2)\,.
\end{equation}
It is straightforward if tedious to compute the Taylor expansions
of the denominator and numerator: we get
\begin{equation} \label{Dexp}
D = -2\kappa^2 a^4 \left( \kappa \!+\! 2\beta^{-1}\right) +
\kO(a^5)\,,
\end{equation}
while in the other expression all the terms cancel up to the third
order giving
\begin{equation} \label{Nexp}
N = 4\kappa^4 a^4 + \kO(a^5)\,.
\end{equation}
The sought kernel is thus
\begin{equation}\label{lim kern}
\sum_{j,j'= -1,0,1} [\Gamma_a(k)]_{jj'}^{-1}\, G_k(x\!-\!y_j)
G_k(x'\!-\!y_{j'}) = -\, {\beta \over 2(2\!+\!\beta\kappa)}\,
e^{-\kappa x} e^{-\kappa x'}\, \left( 1\!+\!\kO(a) \right)
\end{equation}
as expected. In the same way one can treat the other situations
with $x,x'$ belonging to $(-\infty,a),\, (-a,0),\,(0,a)$, and
$(a,\infty)$. In the coefficient this corresponds to different
combinations of $(w,1,w^{-1})$ and $(w^{-1},1,w)$ in
(\ref{sandwich}). Due to the symmetry of
$[{\Gamma_a(i\kappa)}]^{-1}$, however, there are just two
different expressions, the other one having the numerator replaced
by
\begin{equation}\label{N2}
N = (w^4+1)v + 2 [w^2- (1\!+\!u)(1\!+\!v)] + (w^2\!-\!1\!-\!u)
(u\!-\!1\!-\!w^2)
\end{equation}
leading to
\begin{equation}\label{Nexp2}
N = -4\kappa^4 a^4 + \kO(a^5)
\end{equation}
and the correct kernel again; recall the sign factor in
(\ref{signG}). This yields the relation (\ref{resolv approx}).

For a fixed $\kappa>0$ we see from the relation (\ref{sandwich})
that its left-hand side can be majorized by a function from
$L^2(\bR^2)$ which is independent of $a$. The same is, of course,
true for the last term in (\ref{Krein}). Then by (\ref{Krein}),
(\ref{3delta-res}), (\ref{dif}), and dominated convergence the
resolvent converges in the Hilbert-Schmidt norm,
\begin{equation}\label{HSconv}
\lim_{a\rightarrow 0}\,
\left\|(-\Delta_{{\mathcal A}_{a(\epsilon)},Y_{a(\epsilon)}}
+\kappa^2)^{-1}-(\Xi_{\beta,y} + \kappa^2)^{-1} \right\|_2 = 0 \,,
\end{equation}
and thus, {\em a fortiori}, $\{-\Delta_{{\mathcal A}_a,Y_a}\}_{a\geq0}$
approximates $\Xi_{\beta,y}$ in the norm-resolvent topology.
\hfill$\Box$
\begin{rem} 
{\rm The result remains valid if the coupling constants
${\mathcal{A}}_a$ are replaced by
\begin{equation} \label{higher A}
\alpha_{\pm 1}(a) = {2\over \beta} - {1\over a} +\varphi_1(a)\,,
\quad \alpha_0(a) = {\beta\over a^2} \left( 1\!+\!\varphi_0(a)
\right)\,,
\end{equation}
where $\varphi_j$ are smooth functions behaving as
${\mathcal{O}}(a)$ for $a\to 0+\:$.}
\end{rem}


%
\section{Approximation of $\delta'$ by regular potentials}
\setcounter{equation}{0}

It is easy to use the above result to prove the existence of an
approximation of $\delta'$ by local potentials. After a suitable
translation we can put $y = 0$ and we seek in the form
\begin{equation} 
W^a_{\epsilon,0}(x) = \frac{\beta}{\epsilon a(\epsilon)^2}\,
V_0\left(x\over\epsilon\right) + \left(\frac{2}{\beta} -
\frac{1}{a(\epsilon)}\right)\left\{ \frac{1}{\epsilon}\, V_{-1}
\left(\frac{x+a(\epsilon)}{\epsilon}\right) + \frac{1}{\epsilon}\,
V_1 \left(\frac{x-a(\epsilon)}{\epsilon}\right)\right\};
\end{equation}
the general potential approximation $W^a_{\epsilon,y}(x)$ is
obtained by replacing $x$ by $x\!-\!y$ at the right-hand side. In
this expression $\beta \in \bR\setminus \{0\}$, and the involved
potentials are supposed to satisfy $V_j \in L^1(\bR)$ and
\begin{equation} \label{normal}
\int_{\bR} V_j(x)\,dx = 1
\end{equation}
for $j = -1,0,1$. The function $a :\bR_+ \rightarrow \bR_+$, to be
specified later, is supposed to be continuous at $\epsilon = 0$
with $a(0) = 0$. The family of one-dimensional Schr\"odinger
operator used to approximate $\Xi_{\beta,y}$ will be of the form
\begin{equation} \label{approx H}
H^a_{\epsilon,y}:= -\Delta + W^a_{\epsilon,y} \,.
\end{equation}
If $V_j \in L^1(\bR)$ the r.h.s. is defined in the sense of the
corresponding quadratic forms. If we add the requirement $V_j \in
L^2(\bR)$, then $W^a_{\epsilon,y}(x)$ is an infinitely small
perturbation of the Laplacian and (\ref{approx H}) as a
self-adjoint operator is defined on $\dom(H^a_{\epsilon,0})=
\dom(-\Delta) = W^{2,2}(\bR)$ as an operator sum. We will make
this assumption everywhere in the following, except for
Theorem~\ref{reg approx} where we refer directly to a result in
\cite{AGHH}.

To compare the resolvents, we choose $k=i\kappa$ which belongs to
the resolvent sets of both $H^a_{\epsilon,y}$ and the operator
$\Xi_{\beta,y}$ introduced above; this can be achieved if $k^2$ is
nonreal or with $\kappa>0$ large enough. Then we may employ the
elementary estimate
\begin{eqnarray}\label{est}
 \lefteqn{ \left\|(H^a_{\epsilon,y} \!+\!\kappa^2)^{-1} - ( \Xi_{\beta,y} \!+\!
\kappa^2)^{-1}\right\| } \\ && \leq \left\|(H^a _{\epsilon,y}
\!+\!\kappa^2)^{-1} - (-\Delta_{{\mathcal
A}_{a(\epsilon)},Y_{a(\epsilon)}}\!+\!\kappa^2)^{-1}\right\| +
\left\|(-\Delta_{{\mathcal
A}_{a(\epsilon)},Y_{a(\epsilon)}}\!+\!\kappa^2)^{-1}-(\Xi_{\beta,y}
\!+\! \kappa^2)^{-1}\right\| \nonumber
\end{eqnarray}
to prove the following claim:
\begin{theo} \label{reg approx}%
Let $V_j \in L^1(\bR),\: j=-1,0,1$. For any sequence
$\{a_n\}\subset (0,\infty)$ with $a_n\to 0$ there is a sequence
$\{\epsilon_n\}$ of positive numbers with $\epsilon_n\to 0$ such
that
\begin{equation} \label{pot approx}
\lim_{n\to\infty}\, \left\|(H^{a_n}_{\epsilon_n,y} +\kappa^2)^{-1}
- (\Xi_{\beta,y} + \kappa^2)^{-1}\right\| = 0
\end{equation}
holds for any $\kappa > 2|\beta|^{-1}$.
\end{theo}
{\em Proof:} Without loss of generality we may put $y=0$. In view
of Theorem~\ref{delta approx} it is sufficient to deal with the
first term at the right-hand side of (\ref{est}). By
\cite[Thm.~II.2.2.2]{AGHH} for each $a_n >0$, $n = 1,2,\ldots$,
there exists a sequence of $\{\epsilon_{nm}\}^{\infty}_{m=1}$ with
$\lim_{m\to\infty}\epsilon_{nm}= 0$ such that
\begin{equation}\label{delta ap}
\lim_{m\to 0}\, \left\|(H^{a_n}_{\epsilon_{nm},0} +\kappa^2)^{-1}
- (-\Delta_{{\mathcal A}_{a_n},Y_{a_n}}+\kappa^2)^{-1}\right\| = 0 \,,
\end{equation}
where $Y_{a_n} = \{y_j\}_{j=-1,0,1} = \{-a_n,0,a_n\}$,
$A_{a_n}= \{\alpha_{j}^{(n)}\}_{j=-1,0,1} =
(2\beta^{-1} - a_n^{-1},\beta a_{n}^{-2}, 2\beta^{-1} -
a_n^{-1})$ and $\{H^{a_n}_{\epsilon_{nm}, 0}\}_{n\geq1}$ are defined by local
potentials
\begin{eqnarray}\label{pot ap2}
\lefteqn{W^{a_n}_{\epsilon_{nm},0}(x) =
\frac{\beta}{{\epsilon_{nm}}\, a_{n}^2}V_0
\left(x\over{\epsilon_{nm}} \right) }\\ & &
 \phantom{AAAA} + \left(\frac{2}{\beta} - \frac{1}{a_n}\right)\left\{
 \frac{1}{\epsilon_{nm}}\, V_{-1}\left(\frac{x+a_n}{\epsilon_{nm}}\right)
 + \frac{1}{\epsilon_{nm}}\,
 V_1\left(\frac{x-a_n}{\epsilon_{nm}}\right)\right\} \,.
\nonumber
\end{eqnarray}
Indeed, in view of (\ref{normal}) Theorem II.2.2.2 of \cite{AGHH}
applies if we choose the real analytic function $\lambda_j(\cdot)$,
which enters into Theorem II.2.2.2, of the form
$\lambda_j(\epsilon_{nm}) := \epsilon_{nm}\alpha^{(n)}_j$.
If $\imag k^2\ne 0$ the norms at the right-hand side
of (\ref{delta ap}) are uniformly bounded and the claim is valid
for the diagonal sequence, $\epsilon_n: = \epsilon_{nn}$ --
cf.~\cite[Sec.~I.3]{RS}. By the first resolvent identity its
validity extends to any point outside the spectrum of
$\Xi_{\beta,0}$. \hfill$\Box$ \vspace{2mm}

The diagonal trick used in the above proof introduces a relation
between the parameters $a$ and $\epsilon$. Since to a given $a$ we
choose $\epsilon$ small enough to meet the requirements, the
procedure works if $a(\epsilon)$ tends to zero sufficiently slowly
as $\epsilon\to 0+$. Put like that the claim is, of course, very
vague. Even without computing the resolvents, e.g., we can
conjecture that the family (\ref{approx H}) will {\em not} yield
the sought approximation if $a(\epsilon)\sim \epsilon^{\nu}$ with
$\nu>1$ since then the three potentials will overlap substantially
for small values of $\epsilon$ and eventually the (divergent)
overall mean value will prevail.

The question about a rate between $a$ and $\epsilon$ which is
sufficient to yield a convergent approximation is subtle, and the
rest of the section is devoted to it. As above we put $y=0$ in the
following argument restoring a general $y$ only in the final
result. First we introduce the sesquilinear forms
$t^{(0)}_{a,\epsilon}[\cdot,\cdot]$,
$$ 
t^{(0)}_{a,\epsilon}[u,v] :=
\frac{\beta}{a^2}\left\{u(0)\overline{v(0)} -
\frac{1}{\epsilon}\int^{+\infty}_{-\infty} dx\;
V_0(x/\epsilon)u(x)\overline{v(x)}\right\}\,, $$
and $t^{(j)}_{a,\epsilon}[\cdot,\cdot]$,
$$ 
t^{(j)}_{a,\epsilon}[u,v] := \left(\frac{2}{\beta} -
\frac{1}{a}\right)\left\{u(ja)\overline{v(ja)} -
\frac{1}{\epsilon}\int^{+\infty}_{-\infty} dx \; V_j(x -
ja/\epsilon)u(x)\overline{v(x)}\right\}\,, $$
where $j = \pm 1$ and $\dom(t^{(0)}_{a,\epsilon}) =
\dom(t^{(j)}_{a,\epsilon}) = W^{1,2}(\bR)$. We set
$$ 
t_{a,\epsilon}[\cdot,\cdot] := t^{(0)}_{a,\epsilon}[\cdot,\cdot] +
t^{(-1)}_{a,\epsilon}[\cdot,\cdot] +
t^{(+1)}_{a,\epsilon}[\cdot,\cdot] $$
with $\dom(t_{a,\epsilon}) = W^{1,2}(\bR)$. To proceed further we
need stronger hypotheses about the potentials, namely the
conditions (\ref{cond-0}) and (\ref{cond-j}) below. It can be shown
that in combination with $V_j \in L^2(\bR)$ they imply $V_j \in
L^1(\bR)$.

\begin{lem} \label{estimate-0}
Let $V_0 \in L^2(\bR)$. If the conditions (\ref{normal}) and
\begin{equation} \label{cond-0}
\int^{+\infty}_{-\infty} dx \; |x|^{1/2}\;|V_0(x)| < +\infty\,,
\end{equation}
are valid, then $|t^{(0)}_{a,\epsilon}[u,v]| \le \sqrt{2}\sqrt{\epsilon}|\beta|
a^{-2}\int^{+\infty}_{-\infty} dx \; |x|^{1/2}\;|V_0(x)| \;
\|u\|_{W^{1,2}}\|v\|_{W^{1,2}} $ holds for $u,v \in W^{1,2}(\bR)$.
\end{lem}
{\em Proof:} Changing the integration variable $x\to\epsilon x$ in
the definition of $t^{(0)}_{a,\epsilon}[u,v]$ we get
$$ 
t^{(0)}_{a,\epsilon}[u,v] =
\frac{\beta}{a^2}\int^{+\infty}_{-\infty} dx \;
V_0(x)\left\{u(0)\overline{v(0)} - u(\epsilon
x)\overline{v(\epsilon x)}\right\}\,, $$
which yields
$$ 
t^{(0)}_{a,\epsilon}[u,v] =
-\frac{\beta}{a^2}\int^{+\infty}_{-\infty} dx \; V_0(x)
\left\{(u(0) - u(\epsilon x))\overline{v(0)} +
u(\epsilon x)\overline{(v(0) - v(\epsilon x))}\right\}.
$$
Since
\begin{equation}\label{est4}
|f(x)| \le \frac{1}{\sqrt{2}}\|f\|_{W^{1,2}}, \quad f \in
W^{1,2}(\bR),
\end{equation}
and
\begin{equation}\label{est5}
|f(x) - f(y)| \le \sqrt{|x-y|} \, \|f\|_{W^{1,2}}, \quad f \in W^{1,2}(\bR),
\end{equation}
as it follows from $f(x)-f(y)= -\int_x^y f'(t)\,dt$ and the
H\"older inequality, we find
$$ 
|t^{(0)}_{a,\epsilon}[u,v]| \le
2\sqrt{\frac{\epsilon}{2}} \, \frac{|\beta|}{a^2}\int^{+\infty}_{-\infty} dx \,
\sqrt{|x|}\,|V_0(x)| \; \|u\|_{W^{1,2}}\|v\|_{W^{1,2}}
$$
for $u,v \in W^{1,2}(\bR)$ which proves the lemma.\hfill$\Box$ \vspace{2mm}

\begin{lem}\label{estimate-j}
Let $V_j \in L^2(\bR)$, $j = \pm 1$, and $\beta \not= 0$. If the
conditions (\ref{normal})  and
\begin{equation} \label{cond-j}
\int^{+\infty}_{-\infty} dx \; |x|^{1/2}\; |V_j(x)| < +\infty\,, \quad j
= \pm 1\,,
\end{equation}
are satisfied, then
\begin{equation} \label{est11}
\left|t^{(j)}_{a,\epsilon}[u,v]\right| \le \sqrt{2}\sqrt{\epsilon}
\,\left|\frac{2}{\beta} - \frac{1}{a}\right|
\int^{+\infty}_{-\infty} dx \; |x|^{1/2}\; |V_j(x)| \;
\|u\|_{W^{1,2}}\|v\|_{W^{1,2}}
\end{equation}
holds for any $u,v \in  W^{1,2}(\bR)$ and $j = \pm 1$.
\end{lem}
{\em Proof:} Let $j = -1$. Changing the integration variable to
$\epsilon x -a$ in the definition of $t^{(-1)}_{a,\epsilon}[u,v]$
we get
$$ 
t^{(-1)}_{a,\epsilon}[u,v] = \left(\frac{2}{\beta} -
\frac{1}{a}\right) \int^{+\infty}_{-\infty} dx \; V_{-1}(x)
\left\{u(-a)\overline{v(-a)} - u(\epsilon x -a)
\overline{v(\epsilon x - a )}\right\}. $$
From here we infer
\begin{eqnarray}
\lefteqn{t^{(-1)}_{a,\epsilon}[u,v] = \left(\frac{2}{\beta} - \frac{1}{a}\right)\times}\\
& & \hspace{-0.5cm}
\int^{+\infty}_{-\infty} dx \; V_{-1}(x)
\left\{(u(-a) - u(\epsilon x -a))\overline{v(-a)} +
u(\epsilon x -a)\overline{(v(-a) - v(\epsilon x - a ))}\right\}.\nonumber
\end{eqnarray}
Using again (\ref{est4}) and (\ref{est5}) we complete the proof.
\hfill$\Box$ \vspace{2mm}

\begin{cor}\label{estmate}
Let $V_j \in L^2(\bR)$, $j = -1,0,+1$,  and $\beta \not= 0$. If the potentials
$V_j$ satisfy the conditions (\ref{normal}), (\ref{cond-0}),
(\ref{cond-j}), then the estimate
$$ 
\left|t_{a,\epsilon}[u,v]\right| \le \sqrt{\epsilon}\,
C(a)\;\|u\|_{W^{1,2}}\|v\|_{W^{1,2}}
$$
is valid for $u,v \in \dom(t_{a,\epsilon}) = W^{1,2}(\bR)$, where
the constant $C(a)$ is given by
\begin{eqnarray}\label{est19}
\lefteqn{C(a) :=
\sqrt{2}\Bigg\{\frac{|\beta|}{a^2}\int^{+\infty}_{-\infty} dx \;
|x|^{1/2}\;|V_0(x)|} \nonumber \\ & & \hspace{-0.5cm}
\phantom{AAA} + \left|\frac{2}{\beta} - \frac{1}{a}\right|
\int^{+\infty}_{-\infty} dx \; |x|^{1/2}\{|V_{-1}(x)| +
|V_{+1}(x)|\}\Bigg\}.
\end{eqnarray}
\end{cor} \vspace{3mm}

\noindent Let us next
introduce the operator $G(a): L^2(\bR) \to \bC^3$,
$$ 
G(a)f := \left(\begin{array}{c}
\int^{+\infty}_{-\infty} dx \; G_{i\kappa}(x + a)f(x)\\[2mm]
\int^{+\infty}_{-\infty} dx \; G_{i\kappa}(x)f(x)\\[2mm]
\int^{+\infty}_{-\infty} dx \; G_{i\kappa}(x - a)f(x)
\end{array}\right)
$$
for $f \in \dom(G(a)) = L^2(\bR)$. Obviously, the action of the
adjoint operator $G(a)^*: \bC^3 \to L^2(\bR)$ is given by
$$ 
\left(G(a)^*\xi
%
%
%
%
\right)(x) =
G_{i\kappa}(x + a)\xi_{-1} + G_{i\kappa}(x)\xi_0 + G_{i\kappa}(x
- a)\xi_{+1}\,, $$
where
$$  
\xi := \left(\begin{array}{c}
\xi_{-1}\\
\xi_0\\
\xi_{+1}
\end{array}\right) \in \bC^3.
$$
With these definitions the r.h.s. of (\ref{3delta-res}) can
rewritten as
\begin{equation} \label{fact3}
( -\Delta_{{\mathcal A}_a,Y_a} +
\kappa^2)^{-1}f = (H_0 + \kappa^2)^{-1}f +
G(a)^*\Gamma_a(i\kappa)^{-1}G(a)f\,,
\end{equation}
where $Y_a = \{y_j\}_{j = -1,0,+1}$ with $y_j = ja$ and the matrix
$\Gamma_a(i\kappa)$ is given by (\ref{Gamma}). Furthermore, we
introduce the operator $\hat{G}(a)$:
\begin{equation} \label{fact4}
\hat{G}(a)f := \left(\begin{array}{c}
(H_0 + \kappa^2)^{-1/2}f\\
G(a)f
\end{array}\right) : L^2(\bR) \longrightarrow
\begin{array}{c}
L^2(\bR)\\
\oplus\\
\bC^3
\end{array}
\end{equation}
and the operator $\hat{\Gamma}_a(i\kappa)$:
\begin{equation} \label{fact5}
\hat{\Gamma}_a(i\kappa) := \left(\begin{array}{cc}
I & 0\\
0 & \Gamma_a(i\kappa)
\end{array}\right): \begin{array}{c}
L^2(\bR)\\
\oplus\\
\bC^3
\end{array} \longrightarrow \begin{array}{c}
L^2(\bR)\\
\oplus\\
\bC^3
\end{array}
\end{equation}
Using the definitions (\ref{fact4}) and (\ref{fact5}) we can
rewrite (\ref{fact3}) as
$$ 
( -\Delta_{{\mathcal A}_a,Y_a} + \kappa^2)^{-1}f =
\hat{G}(a)^*\hat{\Gamma}_a(i\kappa)^{-1}\hat{G}(a)f\,. $$
Since $G_{i\kappa}(x - ja) \in W^{1,2}(\bR)$ for $j = -1,0,+1$,
one gets that $\ran(G(a)^*) \subseteq  W^{1,2}(\bR)$, and
consequently, $\ran(\hat{G}(a)^*) \subseteq  W^{1,2}(\bR)$. Thus
it makes sense to define the following sesqui\-linear form
$$ 
d_{a,\epsilon}[\hat{\xi},\hat{\eta}] :=
t_{a,\epsilon}[\hat{G}(a)^*\hat{\xi},\hat{G}(a)^*\hat{\eta}],
\quad \hat{\xi},\hat{\eta} \in \dom(d_{a,\epsilon}) = \hat{\kH} :=
\begin{array}{c}
L^2(\bR)\\
\oplus\\
\bC^3
\end{array}
$$
where
$$ 
\hat{\xi} := \left(\begin{array}{c}
f\\
\xi
\end{array}\right) \quad \mbox{and} \quad \hat{y}  := \left(\begin{array}{c}
g\\
\eta
\end{array}\right)
$$
with $f,g \in L^2(\bR)$ and $\xi,\eta \in \bC^3$. By construction,
the form $d_{a,\epsilon}[\cdot,\cdot]$ defines a bounded operator
$D_{a,\epsilon}: \hat{\kH} \to \hat{\kH}$.

\begin{lem}\label{diff}
Let $V_j \in L^2(\bR)$, $j = -1,0,+1$, and $\beta \not= 0$. If the potentials
$V_j$ satisfy the conditions (\ref{normal}), (\ref{cond-0}) and
(\ref{cond-j}) and $\kappa \ge 1$, then one has
\begin{equation} \label{fact9}
\|D_{a,\epsilon}\|_{\kB(\hat{\kH},\hat{\kH})} \le 4\sqrt{\epsilon}\, C(a)
\end{equation}
for $a > 0$.
\end{lem}
{\em Proof:} Using Corollary~\ref{estmate} we find
$$ 
|d_{a,\epsilon}[\hat{\xi},\hat{\eta}]| =
|t_{a,\epsilon}[\hat{G}(a)^*\hat{\xi},\hat{G}(a)^*\hat{\eta}]| \le
\sqrt{\epsilon}\, C(a)\|\hat{G}(a)^*\hat{\xi}\|_{W^{1,2}}
\;\|\hat{G}(a)^*\hat{\eta}]\|_{W^{1,2}}\,. $$
Since
$$ 
(\hat{G}(a)^*\hat{\xi})(x) = (H_0 + \kappa^2)^{-1/2}f +
G_{i\kappa}(x + a)\xi_{-1} + G_{i\kappa}(x)\xi_0 +
G_{i\kappa}(x-a)\xi_{+1}\,, $$
we have
\begin{eqnarray*} 
\lefteqn{\|\hat{G}(a)^*\hat{\xi}\|^2_{W^{1,2}}  \le 4\Big(\|(H_0 +
\kappa^2)^{-1/2}f\|^2_{W^{1,2}} + \|G_{i\kappa}(\cdot +
a)\xi_{-1}\|^2_{W^{1,2}} }\\ && \phantom{AAAAAAA} +
\|G_{i\kappa}(\cdot)\xi_0\|^2_{W^{1,2}} + \|G_{i\kappa}(\cdot
-a)\xi_{+1}\|^2_{W^{1,2}}\Big) \phantom{AAAAAA} 
\end{eqnarray*}
The assumption $\kappa \ge 1$ yields
$$ 
\|(H_0 + \kappa^2)^{-1/2}f\|^2_{W^{1,2}} \le \|f\|^2, \quad f \in
L^2(\bR)\,, $$
and
$$ 
\|G_{i\kappa}(\cdot - ja)\xi_{j}\|^2_{W^{1,2}} = \frac{1}{4}\left(\kappa^{-1} +
\kappa^{-3}\right)|\xi_j|^2 \le |\xi_j|^2, \quad j = -1,0,+1\,, $$
for $a \ge 0$. In this way we get the estimate
$$ 
\|\hat{G}(a)^*\hat{\xi}\|^2_{W^{1,2}} \le 4\left( \|f\|^2 +
\|\xi\|^2_{\bC^3}\right) \le 4\|\hat{\xi}\|^2_{\hat{\kH}}\,. $$
or
for $a \ge 0$. This leads to the estimate
$$ 
|d_{a,\epsilon}[\hat{\xi},\hat{\eta}]| =
|t_{a,\epsilon}[\hat{G}(a)^*\hat{\xi},\hat{G}(a)^*\hat{\eta}]| \le
4 \sqrt{\epsilon}\, C(a)\|\hat{\xi}\|_{\hat{\kH}}
\|\hat{\eta}\|_{\hat{\kH}} $$
from which (\ref{fact9}) follows readily.\hfill$\Box$ \vspace{3mm}

Let us further introduce the Neumann iterations
$R^{(n)}_{a,\epsilon}(i\kappa)$ defined by
$$ 
R^{(n)}_{a,\epsilon}(i\kappa) :=
\hat{G}^*(a)\hat{\Gamma}_a(i\kappa)^{-1}
\left(D_{a,\epsilon}\hat{\Gamma}_a(i\kappa)^{-1}\right)^n\hat{G}(a), \quad n
=0,1,2,\ldots\,. $$
for $k > \max(-2/\beta,1)$ and $a \in (0,a_0(\kappa))$. The
meaning of these expressions will become clear below; we note that
\begin{equation} \label{neu2}
R^{(0)}_{a,\epsilon}(i\kappa) = ( -\Delta_{{\mathcal A}_a,Y_a} +
\kappa^2)^{-1}.
\end{equation}
We also need to know how the norm of $\Gamma_a(i\kappa)^{-1}$
behaves as $a\to 0$. The Taylor expansion for all the expressions
contained in (\ref{inverse}) yields
\begin{eqnarray*}
 \lefteqn{
 [{\Gamma_a(i\kappa)}]^{-1} = \frac{2\beta a^{-2}}
 {2+\beta\kappa} } \\ && \nonumber \\
&&  \times\left(
\begin{array}{ccc}
 2\kappa(\kappa+\beta^{-1}) & -2\kappa(\kappa+2\beta^{-1}) & 2\kappa\beta^{-1} \\
 -2\kappa(\kappa+2\beta^{-1}) & 4\kappa(\kappa+2\beta^{-1}) &
 -2\kappa(\kappa+2\beta^{-1}) \\
 2\kappa\beta^{-1} &  -2\kappa(\kappa+2\beta^{-1}) & 2\kappa(\kappa+\beta^{-1})
\end{array} \right) \left(1+{\mathcal{O}}(a)\right)\,. \nonumber
\end{eqnarray*}
Consequently, for $\kappa > \max(-2/\beta,1)$ there is a constant
$C_\Gamma(\kappa) > 0$ such that
\begin{equation}\label{neu3}
\left\|\Gamma_a(i\kappa)^{-1}\right\|_{\kB(\bC^3,\bC^3)} \le
C_\Gamma(\kappa) \; a^{-2}
\end{equation}
holds for any $a \in (0,a_0(\kappa))$.

\begin{lem}\label{neu}
Let $V_j \in L^2(\bR)$, $j = -1,0,+1$, and $\kappa > \max(-2/\beta,1)$, $\beta
\not= 0$. If the potentials $V_j$ satisfy the conditions
(\ref{normal}), (\ref{cond-0}) and (\ref{cond-j}),
then the Neumann iterations obey the estimate
\begin{equation} \label{neu4}
\left\|R^{(n)}_{a,\epsilon}(i\kappa)\right\|  \le \frac{2C_\Gamma(\kappa)}{a^2}
\left(\frac{4 \sqrt{\epsilon}\, C_\Gamma(\kappa) C(a)}{a^2}\right)^n
\end{equation}
for $a \in (0,a_0(\kappa))$ and $\,n = 1,2,\ldots\,$, where $C(a)$
is given by (\ref{est19}). 
\end{lem}
{\em Proof:} Since $\kappa > 1$, we have
$$ 
\|\hat{G}(a)\|_{\kB(\kH,\hat{\kH})} =
\|\hat{G}^*(a)\|_{\kB(\hat{\kH},\kH)} \le \sqrt{2}\,. $$
An elementary estimate,
$\left\|R^{(n)}_{a,\epsilon}(i\kappa)\right\| \le  2
\left\|\Gamma_a(i\kappa)^{-1}\right\|^{n+1}_{\kB(\bC^3,\bC^3)} \;
\|D_{a,\epsilon}\|^n_{\kB(\hat{\kH},\hat{\kH})}$, gives
$$ 
\left\|R^{(n)}_{a,\epsilon}(i\kappa)\right\| \le 2 \cdot 4^n
\epsilon^{n/2} C_\Gamma(\kappa)^{n+1} C(a)^n a^{-(2n +2)} $$
so (\ref{neu4}) follows readily.\hfill$\Box$ \vspace{3mm}

If $\kappa > \max\{-2/\beta,1\}$ and the condition
\begin{equation} \label{neu8}
\tau(\epsilon,a,\kappa) := \frac{4 \sqrt{\epsilon}\, C_\Gamma(\kappa) C(a)}{a^2} < 1
\end{equation}
is satisfied for some $a \in (0,a_0(\kappa))$, then the operator
$R_{a,\epsilon}(i\kappa)$,
$$ 
R_{a,\epsilon}(i\kappa) :=
\sum^\infty_{n=0}R^{(n)}_{a,\epsilon}(i\kappa)\,, $$
is well defined. We denote the
closed quadratic form which is associated with the self-adjoint
operator $H^a_{\epsilon,0}$ by $h^a_{\epsilon,0}[\cdot,\cdot]$.
Obviously, its domain is $\dom(h^a_{\epsilon,0}) = W^{1,2}(\bR)$;
we note that the natural norm $\|\cdot\|_{h^a_{\epsilon,0}}$ on
$\dom(h^a_{\epsilon,0})$ is equivalent to the norm of the Hilbert
space $W^{1,2}(\bR)$.

\begin{lem}\label{exist}
Let $V_j \in L^2(\bR)$, $j = -1,0,+1$,  and $\kappa > \max(-2/\beta,1)$, $\beta
\not= 0$. If the potentials $V_j$ satisfy the conditions
(\ref{normal}), (\ref{cond-0}) and (\ref{cond-j}) and
$\tau(\epsilon,a,\kappa) < 1$ is valid for some $a
\in (0,a_0(\kappa))$, then $-\kappa^2$ belongs to the resolvent
set of the operator $H^a_{\epsilon,0}$ given by (\ref{approx H}),
and moreover, one has
\begin{equation} \label{exist1}
(H^a_{\epsilon,0} + \kappa^2)^{-1} = R_{a,\epsilon}(i\kappa)\,.
\end{equation}
\end{lem}
{\em Proof:} Combining the above definitions of the quadratic
forms, we get
\begin{equation} \label{exist2}
\left(\sqrt{-\Delta_{{\mathcal A}_a,Y_a}
+\kappa^2}\;u,\sqrt{-\Delta_{{\mathcal A}_a,Y_a} + \kappa^2}\;v\right) =
h^a_{\epsilon,0}[u,v] + \kappa^2(u,v) + t_{a,\epsilon}[u,v]
\end{equation}
for $u,v \in W^{1,2}(\bR)$. We use this relation for $u =
R_{a,\epsilon}(i\kappa)f$ and $v = \left(-\Delta_{{\mathcal A}_a,Y_a}
+\kappa^2\right)^{-1}g$ with $f,g \in L^2(\bR)$. Since
\begin{equation} \label{exist3}
R_{a,\epsilon}(i\kappa) =
\hat{G}(a)^*\hat{\Gamma}_a(i\kappa)^{-1}\sum^\infty_{n=0}
\left(D_{a,\epsilon}\hat{\Gamma}_a(i\kappa)^{-1}\right)^n\hat{G}(a)
\end{equation}
and $\ran(\hat{G}(a)^*) \subseteq W^{1,2}(\bR)$ we get $u \in
W^{1,2}(\bR)$. Since $v = \left(-\Delta_{{\mathcal A}_a,Y_a}
+\kappa^2\right)^{-1}g \in W^{1,2}(\bR)$, we can insert $u$ and
$v$ into (\ref{exist2}). This yields
\begin{eqnarray*} 
\lefteqn{(R_{a,\epsilon}(i\kappa)f,g) =
h^a_{\epsilon,0}[R_{a,\epsilon}(i\kappa)f,(-\Delta_{{\mathcal A}_a,Y_a}
+\kappa^2)^{-1}g] }\\ & & \phantom{AAAAAA}
+\kappa^2(R_{a,\epsilon}(i\kappa)f,(-\Delta_{{\mathcal A}_a,Y_a}
+\kappa^2)^{-1}g) +
t_{a,\epsilon}[R_{a,\epsilon}(i\kappa)f,(-\Delta_{{\mathcal A}_a,Y_a}
+\kappa^2)^{-1}g]\,. 
\end{eqnarray*}
Using (\ref{neu2}) and (\ref{exist3}) we find
\begin{eqnarray*} 
\lefteqn{t_{a,\epsilon}[R_{a,\epsilon}(i\kappa)f,(-\Delta_{{\mathcal A}_a,Y_a}
+\kappa^2)^{-1}g] }\\ & = &
t_{a,\epsilon}[\hat{G}(a)^*\hat{\Gamma}_a(i\kappa)^{-1}\sum^\infty_{n=0}
\left(D_{a,\epsilon}\hat{\Gamma}_a(i\kappa)^{-1}\right)^n\hat{G}(a)
f,\hat{G}(a)^*\hat{\Gamma}(i\kappa)^{-1}\hat{G}(a) g] 
\\ & = &
\left(D_{a,\epsilon}\hat{\Gamma}_a(i\kappa)^{-1}\sum^\infty_{n=0}
\left(D_{a,\epsilon}\hat{\Gamma}_a(i\kappa)^{-1}\right)^n\hat{G}(a)
f,\hat{\Gamma}(i\kappa)^{-1}\hat{G}(a) g\right)\nonumber\\ & = &
\left(\sum^\infty_{n=1}R^{(n)}_{a,\epsilon}(i\kappa)f,g\right)\,.
\end{eqnarray*}
Furthermore, from (\ref{neu2}) we infer that
\begin{eqnarray*} 
\lefteqn{\left((-\Delta_{{\mathcal A}_a,Y_a} +\kappa^2)^{-1}f,g\right) }\\ &
& = h^a_{\epsilon,0}[R_{a,\epsilon}(i\kappa)f,(-\Delta_{{\mathcal A}_a,Y_a}
+\kappa^2)^{-1}g] +
\kappa^2(R_{a,\epsilon}(i\kappa)f,(-\Delta_{{\mathcal A}_a,Y_a}
+\kappa^2)^{-1}g)\,. 
\end{eqnarray*}
Setting now $h := (-\Delta_{{\mathcal A}_a,Y_a} +\kappa^2)^{-1}g$ we find
\begin{equation} \label{exist7}
(f,h) = h^a_{\epsilon,0}[R_{a,\epsilon}(i\kappa)f,h] +
\kappa^2(R_{a,\epsilon}(i\kappa)f,h)
\end{equation}
for $h \in \dom(-\Delta_{{\mathcal A}_a,Y_a})$. Since
$\dom(-\Delta_{{\mathcal A}_a,Y_a})$ is a core for the quadratic form
$h^a_{\epsilon,0}[\cdot,\cdot]$ one concludes that the equality
(\ref{exist7}) extends to each $h \in \dom( h^a_{\epsilon,0})$.
In particular, if $h \in \dom(H^a_{\epsilon,0})$ we have
$$ 
(f,h) = (R_{a,\epsilon}(i\kappa)f,(H^a_{\epsilon,0} +
\kappa^2)h)\,. $$
In this way we find $R_{a,\epsilon}(i\kappa)f \in
\dom(H^a_{\epsilon,0}))$ and
$$ 
(H^a_{\epsilon,0} + \kappa^2)R_{a,\epsilon}(i\kappa)f = f, \quad f
\in \kH\,, $$
and
$$ 
R_{a,\epsilon}(i\kappa)(H^a_{\epsilon,0} + \kappa^2)h = h, \quad h
\in \dom(H^a_{\epsilon,0})\,. $$
Hence $\ker(H^a_{\epsilon,0} + \kappa^2) = \{0\}$ and
$\ran(H^a_{\epsilon,0} + \kappa^2) = \kH$, so the operator
$H^a_{\epsilon,0} + \kappa^2$ is boundedly invertible and
$(H^a_{\epsilon,0} + \kappa^2)^{-1} = R_{a,\epsilon}(i\kappa)$.
\hfill$\Box$ \vspace{3mm}

With the help of Lemma \ref{exist} one can prove the following
estimate.
\begin{lem}\label{diff0}
Let $V_j \in L^2(\bR)$, $j = -1,0,+1$, and $\kappa > \max(-2/\beta,1)$, $\beta
\not= 0$. If the potentials $V_j$ satisfy the conditions
(\ref{normal}), (\ref{cond-0}) and (\ref{cond-j}) and
$\tau(\epsilon,a,\kappa) < 1$ is valid for some $a \in (0,a_0(\kappa))$, then
\begin{equation} \label{dif1}
\left\|(H^a_{\epsilon,0} + \kappa^2)^{-1} - (-\Delta_{{\mathcal A}_a,Y_a}
+\kappa^2)^{-1}\right\| \le
2C_\Gamma(\kappa) \;
\frac{\tau(\epsilon,a,\kappa)}{a^2} \left(1- \tau(\epsilon,a,\kappa)
\right)^{-1}.
\nonumber
\end{equation}
\end{lem}
{\em Proof:} Taking into account (\ref{exist1}) and (\ref{neu2})
we find
$$ 
(H^a_{\epsilon,0} + \kappa^2)^{-1} - (-\Delta_{{\mathcal A}_a,Y_a} +
\kappa^2)^{-1} = \sum^\infty_{n=1}
R^{(n)}_{a,\epsilon}(i\kappa)\,. $$
Using the notation (\ref{neu8}) and taking into account
the estimate (\ref{neu4}) one gets
$$ 
\left\|(H^a_{\epsilon,0} + \kappa^2)^{-1} - (-\Delta_{{\mathcal A}_a,Y_a} +
\kappa^2)^{-1}\right\| \le \frac{2 C_\Gamma(\kappa)}{a^2}
\sum^\infty_{n=1} \tau(\epsilon,a,\kappa)^n. $$
If $\tau(\epsilon,a,\kappa) < 1$ is satisfied, we obtain
(\ref{dif1}) easily. \hfill$\Box$\vspace{2mm}

Now we are ready to say something about the rate of the potential
approximation in terms of the relation between $a$ and $\epsilon$.
Consider a function $a: (0,\infty) \to (0,\infty)$.

\begin{theo}\label{dif5}
Let $V_j \in L^2(\bR)$, $j = -1,0+1$, and $\kappa >
\max(-2/\beta,1)$, $\beta \not= 0$. Moreover, suppose that
$a(\epsilon)\to 0$ as $\epsilon \to 0+$. If the potentials $V_j$
satisfy the conditions (\ref{normal}), (\ref{cond-0}) and
(\ref{cond-j}) for $j = -1,0,+1$, and
\begin{equation} \label{dif6}
\lim_{\epsilon \to 0} \frac{\epsilon}{a(\epsilon)^{12}} = 0\,,
\end{equation}
then
\begin{equation}\label{dif7}
\lim_{\epsilon \to 0}\left\|(H^a_{\epsilon,y} + \kappa^2)^{-1} -
(-\Delta_{{\mathcal A}_{a(\epsilon)},Y_{a(\epsilon)}} +
\kappa^2)^{-1}\right\| = 0\,
\end{equation}
and
\begin{equation}\label{dif8}
\lim_{\epsilon \to 0}\left\|(H^a_{\epsilon,y} + \kappa^2)^{-1} -
(\Xi_{\beta,y}  + \kappa^2)^{-1}\right\| = 0\,.
\end{equation}
\end{theo}
{\em Proof:} Since $H^a_{\epsilon,y}$ is unitarily equivalent to
$H^a_{\epsilon,0}$ by translation and the same is true for the
other involved operators, we can again put $y=0$ without loss of
generality. By assumption, $a(\epsilon) \in (0,a_0(\kappa))$ for
$\epsilon$ sufficiently small. Further, we note that there is a
constant $C = C(V_j,\beta)$ such that $C(a) \le Ca^{-2}$ for $a >
0$. Using that we can estimate
$$ \tau(\epsilon,a(\epsilon),\kappa) \le \frac{4\sqrt{\epsilon}\,
C_\Gamma(\kappa)C}{a(\epsilon)^4} = 4 C_\Gamma(\kappa)C
a(\epsilon)^2\frac{\sqrt{\epsilon}}{a(\epsilon)^6}\,, $$
so $\lim_{\epsilon \to 0+} \tau(\epsilon,a(\epsilon),\kappa) = 0$
by (\ref{dif6}) and $\lim_{\epsilon \to 0} a(\epsilon) = 0$.
Hence, $\tau(\epsilon,a(\epsilon),\kappa) < 1$ holds for
$\epsilon$ sufficiently small. Applying Lemma \ref{diff0} we get
$$ \left\|(H^a_{\epsilon,0} + \kappa^2)^{-1} -
(-\Delta_{{\mathcal A}_{a(\epsilon)},Y_{a(\epsilon)}} + \kappa^2)^{-1}\right\|
\le 8 C_\Gamma(\kappa)^2C\frac{\sqrt{\epsilon}}{a(\epsilon)^6}(1 -
\tau(\epsilon,a(\epsilon),\kappa))^{-1}. $$
Taking into account once again the assumption (\ref{dif6}) we
prove (\ref{dif7}). Moreover, using Theorem~\ref{delta approx}
together with the estimate (\ref {est}) we arrive at
(\ref{dif8}).\hfill$\Box$ \vspace{3mm}



%
\section{Exceptional character of the CS approximation}
\setcounter{equation}{0}

In conclusion we want to show that it is sufficient to disbalance
the limiting procedure slightly, say by changing the normalization
(\ref{normal}), and the result will be completely different than
that in Theorem~\ref{dif5}. For simplicity we will consider the
case $y = 0$ only. Denote by $-\Delta_{D,0}$ the Laplace operator
with Dirichlet boundary conditions at the origin, i.e
$$ 
\dom(-\Delta_{D,0}) = \left\{f \in W^{2,2}(\bR_-) \oplus
W^{2,2}(\bR_+): f(0-) = f(0+) = 0\right\} $$
and
$$ 
(-\Delta_{D,0}f)(x) = -\frac{d^2}{dx^2}f(x)\,, \quad f \in
\dom(-\Delta_{D,0})\,. $$
With respect to $L^2(\bR) = L^2(\bR_-) \oplus
L^2(\bR_+)$ the operator $-\Delta_{D,0}$ decomposes into
$$ 
-\Delta_{D,0} = -\Delta^-_{D,0} \oplus -\Delta^+_{D,0} $$
with $\dom(-\Delta^\pm_{D,0}) = \left\{f \in
W^{2,2}(\bR_\pm): f(0\pm) = 0 \right\}$. We note that $\sigma(-\Delta^\pm_{D,0}) =
[0,+\infty)$. The resolvents $(-\Delta^{\pm}_{D,0} +
\kappa^2)^{-1}$ are integral operators with the kernels
$$ 
D^{\pm}_{i\kappa}(x,x') := \left\{\begin{array}{ccl}
\pm\frac{1}{\kappa}\;e^{\mp\kappa x}\;\sinh(\kappa x') &
\quad\dots\quad & \pm x' \in [0,\pm x) \\[2mm]
\pm\frac{1}{\kappa}\;\sinh(\kappa x)\;e^{\mp\kappa x'} &
\quad\dots\quad  & \pm x' \in [\pm x,+\infty)
\end{array}\right.
$$
A straightforward computation shows that
\begin{equation} \label{dir6}
G_{i\kappa}(x,x') = D^-_{i\kappa}(x,x') \oplus D^+_{i\kappa}(x,x')
+ \frac{1}{2\kappa}e^{-\kappa|x|}e^{-\kappa|x'|}. \end{equation}
The indicated modification corresponds to the changed
$-\Delta_{{\mathcal A}_a,Y_a}$ with ${\mathcal A}_a$ replaced by
$\alpha {\mathcal A}_a$,
$$ 
\alpha {\mathcal A}_a := \left\{\alpha(2\beta^{-1} \!- a^{-1}),
\alpha \beta a^{-2},\alpha(2\beta^{-1} \!- a^{-1}) \right\}, $$
where $\alpha,\,\beta \in \bR \setminus \{0\}$. The form
$\kQ_{\alpha{\mathcal A}_a,Y_a}[\cdot,\cdot]$ associated with the operator
$-\Delta_{\alpha{\mathcal A}_a, Y_a}$ is given by
$$ \label{dir8} \kQ_{\alpha{\mathcal A}_a,Y_{a}}[u,v] = (u',v') +
\alpha\frac{\beta}{a^2}\, u(0)\overline{v(0)} +
\alpha\left(\frac{2}{\beta} -
\frac{1}{a}\right)\{u(+a)\overline{v(+a)} +
u(-a)\overline{v(-a)}\}\,, $$
where $u,v \in \dom(\kQ_{\alpha {\mathcal A}_a,Y_{a,\alpha}}) = W^{1,2}(\bR)$, which
means that $\alpha\ne 1$ amounts to a simultaneous change of all
the $\delta$ coupling parameters. The resolvent
$(-\Delta_{\alpha{\mathcal A}_a,Y_a} + \kappa^2)^{-1}$ is again given by
Krein's formula
\begin{equation} \label{dir9}
(-\Delta_{\alpha{\mathcal A}_a,Y_a} + \kappa^2)^{-1}(x,x') =
G_{i\kappa}(x,x') -
\sum_{j,j'=-1,0,+1}[\Gamma_{a,\alpha}(i\kappa)]^{-1}_{jj'}
G_{i\kappa}(x-y_j)G_{i\kappa}(x-y_{j'})\,,
\end{equation}
where
$$ 
\Gamma_{a,\alpha}(i\kappa) =
\frac{1}{2\kappa}\left(\begin{array}{ccc} 1 + \alpha u & w & w^2
\\ w & 1 + \alpha v & w \\ w^2 & w & 1 + \alpha u
\end{array}\right),
$$
i.e., in comparison with (\ref{Gamma}) we have $u \to \alpha u$,
$v \to \alpha v$, while $w$ is preserved.

\begin{lem}\label{spec}
Let $\kappa > 0$. The resolvent $(-\Delta_{\alpha {\mathcal A}_a,Y_a} +
\kappa^2)^{-1}$ exists for sufficiently small $a > 0$ if $\alpha
\not= 1$.
\end{lem}
{\em Proof:} It is sufficient that $-\kappa^2$ is not an
eigenvalue. As in Proposition~\ref{spec approx} this would be true
for $-\Delta_{\alpha {\mathcal A}_a,Y_a} $ if  $\kappa$ satisfies one of the
equations analogous to (\ref{spec1}) and (\ref{spec2}), with
$\kappa$ replaced by $\alpha\kappa$ at the r.h.s. The Taylor
expansion around $a = 0$ shows that this cannot happen unless
$\alpha= 1$. \hfill$\Box$ \vspace{2mm}

In the following we fix $\kappa > 0$, $\alpha \not\in \{0,1\}$,
and $\beta \not= 0$. Then there is $a_0(\kappa) > 0$ such that for
all $a \in (0,a_0(\kappa))$ the resolvent $(-\Delta_{\alpha
{\mathcal A}_a,Y_a} + \kappa^2)^{-1}$ exists.

\begin{theo} \label{dirichlet}
Let $\kappa > 0$, $\alpha \not= 0,1$, and $\beta \not= 0$ be fixed.
Then the relation
\begin{equation} \label{dir11}
\lim_{a\rightarrow 0+}\, \left( -\Delta_{\alpha {\mathcal A}_a,Y_a} +
\kappa^2 \right)^{-1}(x,x')= D^-_{i\kappa}(x,x') \oplus
D^+_{i\kappa}(x,x')
\end{equation}
holds for any $x,x'\in \bR$. Consequently,
$-\Delta_{\alpha {\mathcal A}_a,Y_a}\to -\Delta_{D,0}$ as $a\to 0+$ in the
norm-resolvent sense.
\end{theo}
{\em Proof:}
Considering the case $x,x' \ge a$ and following the line of
reasoning from (\ref{inverse}) to (\ref{Dexp}) we obtain
\begin{equation} \label{dir12}
\sum_{jj'=-1,0+1}[\Gamma_{a,\alpha}(i\kappa)]^{-1}_{jj'}G(x \!-\!
y_j)G(x \!-\! y_{j'}) = \frac{1}{4\kappa^2}e^{-\kappa x}e^{-\kappa
x'} \frac{N_\alpha}{D_\alpha} \end{equation}
with
$$ 
D_\alpha := \frac{(w^2 \!- 1 -\alpha u)[(1 \!+\! \alpha u)(1 \!+\!
\alpha v) - w^2(1 \!-\! \alpha v)]}{2\kappa} $$
and
$$ 
N_\alpha := (w^2 \!+ w^{-2})[w^2 \!- (1 \!+\! \alpha u)(1 \!+\!
\alpha v)] + 2\alpha w^2v + (w^2 \!- 1\! \!- \alpha u)(\alpha u
\!-\! 1\! \!- w^2)\,. $$
If $\alpha \not= 1$, one gets
$$ 
D_\alpha = -2\kappa a^2(1-\alpha) + \kO(a^3) $$
and
\begin{equation} \label{dir16}
N_\alpha = -4\kappa^2 a^2(1-\alpha) + \kO(a^3)\,,
\end{equation}
so the r.h.s. of (\ref{dir12}) equals $2\kappa e^{-\kappa
x}e^{-\kappa x'}(1 + \kO(a))$. Inserting (\ref{dir16}) into
(\ref{dir9}) and using (\ref{dir6}), we find
\begin{equation} 
\lim_{a \to +0} (-\Delta_{\alpha {\mathcal A}_a,Y_a} + \kappa^2)^{-1}(x,x') =
D^+_{i\kappa}(x,x')
\end{equation}
for $x,x' \in [a,+\infty)$.  In the same way one can treat the
other combinations with $x,x'$ belonging to $(-\infty,a]$,
$(-a,0)$, $(0,a)$ and $[a,+\infty)$; doing so we check
(\ref{dir11}) for $x,x' \in \bR$.

Taking into account (\ref{dir6}) and
(\ref{dir9}) one easily verifies that $\left( -\Delta_{\alpha
{\mathcal A}_a,Y_a} + \kappa^2 \right)^{-1}(x,x') - D^-_{i\kappa}(x,x') \oplus
D^+_{i\kappa}(x,x')$ can be majorized by a function from
$L^2(\bR^2)$ which is independent of $a$. By (\ref{dir11})
and the Lebesgue convergence theorem the difference $\left(
-\Delta_{\alpha {\mathcal A}_a,Y_a} +\kappa^2 \right)^{-1} - \left(
-\Delta_{D,0} + \kappa^2 \right)^{-1}$ converges to zero in the
Hilbert-Schmidt norm, so $-\Delta_{\alpha {\mathcal A}_a,Y_a}\to
-\Delta_{D,0}$ as $a\to 0+$ in the norm-resolvent
sense.\hfill$\Box$\vspace{3mm}

Let us introduce the Schr\"odinger operator
$H^a_{\epsilon,0,\alpha}$ defined by
$$ 
H^a_{\epsilon,0,\alpha} := -\Delta + \alpha W^a_{\epsilon,0} $$
for $\alpha \in \bR \setminus \{0\}$ as in the previous section.
It corresponds to rescaling of the original approximation
potential: we have $H^a_{\epsilon,0,\alpha} = H^a_{\epsilon,0}$ if
$\alpha = 1$. The Neumann iterations are now defined by
$$ 
R^{(n)}_{a,\epsilon,\alpha}(i\kappa) :=
\hat{G}^*(a)\hat{\Gamma}_{a,\alpha}(i\kappa)^{-1}
\left(D_{a,\epsilon}\hat{\Gamma}_{a,\alpha}(i\kappa)^{-1}\right)^n\hat{G}(a), \quad n
=0,1,2,\ldots\,. $$
for $k > 1$ and $a \in (0,a_0(\kappa))$ where the definition of
$\hat{\Gamma}_{a,\alpha}(i\kappa)$ is obvious.
We note that for for $\kappa > 1$ and
$\alpha \not=1$ there is a constant $C_{\Gamma_\alpha}(\kappa) > 0$ such
that instead of (\ref{neu3}) one has the estimate
$$\|\Gamma_{a,\alpha}(i\kappa)^{-1}\|_{{\cal B}(\bC^3,\bC^3)} \le
C_{\Gamma_\alpha}(\kappa)a^{-1}$$
for $a \in (0,a_0(\kappa))$.
Lemma \ref{neu} reads now as follows.
\begin{lem}\label{dir20-10}
Let $V_j \in L^2(\bR)$, $j -1,0,+1$, and $\kappa > 1$, $\beta
\not= 0$. If the potentials $V_j$ satisfy the conditions
(\ref{normal}), (\ref{cond-0}) and (\ref{cond-j}),
then the Neumann iterations obey the estimate
$$ \left\|R^{(n)}_{a,\epsilon,\alpha}(i\kappa)\right\| \le
\frac{2C_{\Gamma_\alpha}(\kappa)}{a} \left(\frac{4 \sqrt{\epsilon}\,
C_{\Gamma_\alpha}(\kappa) C(a)}{a}\right)^n $$
for $a \in (0,a_0(\kappa))$ and $\,n = 1,2,\ldots\,$, where $C(a)$
is given by (\ref{est19}).
\end{lem}
The proof is similar to that of Lemma \ref{neu}. In view of Lemma
\ref{dir20-10} one has to modify the parameter
$\tau(\epsilon,a,\kappa)$ to
$$\tau_\alpha(\epsilon,a,\kappa) := \frac{4 \sqrt{\epsilon}\,
C_{\Gamma_\alpha}(\kappa)C(a)}{a}.$$
If $\alpha\tau_\alpha(\epsilon,a,\kappa) < 1$
is satisfied, then the operator $R_{a,\epsilon,\alpha}(i\kappa) :=
\sum^\infty_{n=0} \alpha^n R^{(n)}_{a,\epsilon,\alpha}(i\kappa)$ is well
defined. With obvious modifications Lemma \ref{exist} takes the
following form.
\begin{lem}\label{dir20-3}
Let $V_j \in L^2(\bR)$, $j = -1,0,+1$, and let $\kappa > 1$, $\alpha \not\in
\{0,1\}$, and $\beta \not= 0$. If the the potentials $V_j$ satisfy
the conditions (\ref{normal}), (\ref{cond-0}) and (\ref{cond-j})
and $\alpha\tau_\alpha(\epsilon,a,\kappa) < 1$
is valid for some $a \in (0,a_0(\kappa))$, then $-\kappa^2$ belongs
to the resolvent set of the operator $H^a_{\epsilon,0,\alpha}$, and,
moreover, one has
$$ 
(H^a_{\epsilon,0,\alpha} + \kappa^2)^{-1} =
R_{a,\epsilon,\alpha}(i\kappa)\,. $$
\end{lem}
Lemma \ref{diff0} modifies similarly but we get a slightly stronger result
because the matrix $\Gamma_{a,\alpha}(i\kappa)^{-1}$ is now less
singular for any $\kappa>0$ as $a\to 0$.
\begin{lem}\label{dir20-5}
Under the assumptions of the preceding lemma,
$$ 
\left\|(H^a_{\epsilon,0,\alpha} + \kappa^2)^{-1} -
(-\Delta_{\alpha {\mathcal A}_a,Y_a} +\kappa^2)^{-1}\right\| \le
2\alpha C_{\Gamma_\alpha}(\kappa) \frac{\tau_\alpha(\epsilon,a,\kappa)}{a}
\left(1 - \alpha \tau_\alpha(\epsilon,a,\kappa)\right)^{-1}. $$
\end{lem}
Taking into account Theorem \ref{dirichlet} and Lemmata \ref{dir20-3}, \ref{dir20-5}
we thus prove the following theorem.
\begin{theo}\label{dirichlet2}
Let $V_j \in L^2(\bR)$, $j = -1,0,1$, and let $\kappa > 1$,
$\alpha \not\in \{0,1\}$, and $\beta \not= 0$. Furthermore, let
$\lim_{\epsilon \to 0} a(\epsilon) = 0$. If the potentials $V_j$
satisfy the conditions (\ref{normal}), (\ref{cond-0}) and
(\ref{cond-j}) and
$$  
\lim_{\epsilon \to 0} \frac{\epsilon}{a(\epsilon)^8} = 0\,, $$
then
$$ 
\lim_{\epsilon \to 0}\left\|\left(H^a_{\epsilon,0,\alpha} +
\kappa^2\right)^{-1} - \left(-\Delta_{\alpha {\mathcal A}_{a(\epsilon)},
Y_{a(\epsilon)}} + \kappa^2\right)^{-1}\right\| = 0\,. $$
and
$$
\lim_{\epsilon \to 0}\left\|\left(H^a_{\epsilon,0,\alpha} +
\kappa^2\right)^{-1} - \left(-\Delta_{D,0} +
\kappa^2\right)^{-1}\right\| = 0.
$$
\end{theo}
\vspace{2mm}

\noindent Using a translation, the analogous conclusion can be
made for the family $\{H^a_{\epsilon,y,\alpha}\}$ with the
potential center shifted to a point $y$, which naturally converges
for $\alpha \not\in \{0,1\}$ to the Laplacian with the Dirichlet
decoupling at $y$.


\section*{Acknowledgement}

The authors are grateful for the hospitality in the institutes
where parts of this work were done: P.E. and H.N. in Centre de
Physique Th\'eorique, CNRS, Marseille-Luminy, and H.N. and V.Z. in
Nuclear Physics Institute, AS, \v{R}e\v{z} near Prague. We also
thank the referee for pointing out an error in the first version
of the manuscript. The research was partially supported by the
GAAS Grant A1048101 and the Exchange Agreement No.~7919 between
CNRS and the Czech Academy of Sciences.


\end{document}